\documentclass{article}
\usepackage{amssymb,amsmath,bm,geometry,graphics,graphicx,url,lscape,hyperref}
\usepackage{setspace}
\usepackage{color}

\def\pmatrix{\left(\begin{array}}
\def\endpmatrix{\end{array}\right)}
\def\bfx{{\bm{x}}}
\def\bfy{{\bm{y}}}
\def\bfz{{\bm{z}}}
\def\bfe{{\bm{e}}}
\def\bfzero{{\bm{0}}}
\def\RR{\mathbb{R}}
\newtheorem{rem}{Remark}
\def\si2r2{SI$_2$R$_2$}

\begin{document}
\title{A multi-region variant of the SIR model and its extensions}

\author{Luigi Brugnano\,$^{\rm a}$ \and Felice Iavernaro\,$^{\rm b}$}

\date{ \,$^{\rm a}$ {\small Dipartimento di Matematica e Informatica ``U.\,Dini'', Universit\`a di Firenze, Italy, \,\url{luigi.brugnano@unifi.it}}\\ \,$^{\rm b}$ {\small Dipartimento di Matematica, Universit\`a di Bari, Italy, \, \url{felice.iavernaro@uniba.it}}}

\maketitle

\begin{abstract} In this note, we describe simple generalizations of the basic SIR model for epidemic, in case of a multi-region scenario, to be used for predicting the COVID-19 epidemic spread in Italy.


\end{abstract}

\section{Introduction}\label{intro}  A main concern related to the study of the recent  viral pandemic of coronavirus disease (COVID-19) is, of course, to predict its evolution. Concerning its spread in Italy, at the web-page \cite{BI2020} a basic SIR model has been initially used for this purpose.  Even though this simple model has displayed interesting potentialities in yielding accurate predictions of the diffusion of the disease for the subsequent few days  from the last available monitoring report, it heavily relies  on the assumption of {\em homogeneity} of the spread of epidemic across the country, which is often not the case. Furthermore, it cannot capture the possible interactions between the different areas of the country which, as is well known, have played an important role in the diffusion of the epidemic. 

We attempt to describe these important aspects by considering a {\em multi-region} extension of the basic SIR model, 
to be used for the future forecasts. The basic SIR model is recalled in Section~\ref{SIRmod}, whereas its multi-region extension (mrSIR) is described in Section~\ref{mrSIRmod}.

We also define another extension of the SIR model, called \si2r2, which describes a more refined infection mechanism. Also for this model we propose a multi-region extension (mr\si2r2). These are described in Sections~\ref{SARSmod} and \ref{mrSARSmod}, respectively.

\section{SIR model}\label{SIRmod}
The basic SIR (Susceptible, Infectious, Removed) model for the initial spread of an epidemic disease \cite{KK27} (see also, e.g., \cite[pp.\,376--378]{L1979} or \cite{wiki}) is
\begin{eqnarray}\nonumber
\dot x &=& -\beta x y,\\ \label{sir}
\dot y &=& \beta x y -\gamma y,\\ \nonumber
\dot z &=& \gamma y,
\end{eqnarray}
where:
\begin{itemize}
\item $x$ is the number of {\em susceptible},
\item $y$ is the number of {\em infectictious},
\item $z$ is the number of {\em removed},
\end{itemize}
and
\begin{itemize}
\item $\beta$ is the {\em coefficient of infection},
\item $\gamma$ is the {\em coefficient of removal}.
\end{itemize}

It is assumed that recovered people acquire some immunity \cite{biorXiv} (or are dead), so that they are no more susceptible. 
From (\ref{sir}), one easily realizes that:
\begin{equation}\label{cost}
S(t) = x(t)+y(t)+z(t) \equiv const.
\end{equation}
We have used this simple model to predict, initially, the spread of the COVID-19 disease in Italy \cite{BI2020} but, as is clear from the data available from the Italian Protezione Civile \cite{PC2020}, the spread of the disease is not homogeneous in the various regions of Italy.  Another related question to be faced is the effect in the disease diffusion due to  migration of people from severely contaminated areas to regions not yet affected, which causes the settlement of new transmission clusters. The basic SIR model is not able to properly capture these collateral aspects, causing the predictions to become less accurate in the long temporal range, as the spread of the epidemic was going on. In order to cope with the inhomogeneous nature of the diffusion environment and the interactions between areas with different contamination levels, we propose a {\em multi-region} generalization of the basic SIR model. 

\section{Multi-region SIR (mrSIR) model}\label{mrSIRmod}
In order to generalize the model (\ref{sir}) to a scenario where there are $n$ regions, with a different situation of the epidemic, let us define 
\begin{equation}\label{XYZ}
\bfx = \pmatrix{c} x_1\\ \vdots \\ x_n\endpmatrix, \qquad 
\bfy = \pmatrix{c} y_1\\ \vdots \\ y_n\endpmatrix, \qquad 
\bfz = \pmatrix{c} z_1\\ \vdots \\ z_n\endpmatrix,
\end{equation}
as the vectors of susceptible, infectious, and removed in each region (i.e., $x_i,y_i,z_i$ denote the respective levels in the $i$-th region). In so doing, the basic model (\ref{sir}) now becomes:
\begin{eqnarray}\nonumber
\dot \bfx &=& -B\, \bfx \circ \bfy + R\bfx,\\ \label{mrsir}
\dot \bfy &=& B\, \bfx \circ \bfy -C \bfy + R\bfy,\\ \nonumber
\dot \bfz &=& C \bfy,
\end{eqnarray}
where $\circ$ denotes the Hadamard (i.e., component-wise) product, and
\begin{equation}\label{BCR}
B = \pmatrix{ccc} \beta_1\\ &\ddots\\ &&\beta_n\endpmatrix, \qquad
C = \pmatrix{ccc} \gamma_1\\ &\ddots\\ &&\gamma_n\endpmatrix,\qquad
R = \pmatrix{ccc} \rho_{11} & \dots &\rho_{1n}\\ \vdots & &\vdots\\ \rho_{n1}& \dots &\rho_{nn}\endpmatrix,
\end{equation}
are the matrices with the infection, removal, and {\em migration} coefficients. In particular, $\rho_{ij}$ is the coefficient of migration from region $j$ to region $i$. Concerning matrix $R$, we then assume:
\begin{equation}\label{R}
\rho_{jj}\le0, \qquad \rho_{ij}\ge0, ~ i\ne j, \qquad \sum_{i=1}^n \rho_{ij}=0,\qquad j=1,\dots,n,
\end{equation}
i.e.
\begin{equation}\label{R1}
\rho_{jj} = -\sum_{i\ne j} \rho_{ij},  \qquad j=1,\dots,n.
\end{equation}
Conditions (\ref{R})-(\ref{R1})  can be cast in vector form as
$$\bfe^\top R = \bfzero^\top, \qquad \bfe = (1,\dots,1)^\top, ~\bfzero = (0,\dots,0)^\top\in\RR^n,$$
thus providing the conservation property
\begin{equation}\label{cost1}
S(t) = \sum_{i=1}^n \left[x_i(t)+y_i(t)+z_i(t)\right] \equiv const,
\end{equation}
which is the analogous of (\ref{cost}) for the present model. In componetwise form, (\ref{mrsir}) reads, by taking into account (\ref{R})-(\ref{R1}):
\begin{eqnarray}\nonumber
\dot x_i &=& -\beta_i x_i y_i + \sum_{j=1}^n\rho_{ij} x_j,\\ \label{mrsiri}
\dot y_i &=& \beta_i x_i y_i - \gamma_i y_i + \sum_{j=1}^n\rho_{ij} y_j,\\[2mm] \nonumber
\dot z_i &=& \gamma_i y_i,  \qquad\qquad\qquad\qquad i=1,\dots,n.
\end{eqnarray}
This model has been used to update the forecasts at the web-site \cite{BI2020} for both Italy and Spain.\footnote{For Spain, we have used the data from Spanish  Centro de Coordinaci\'on de Alertas y Emergencias Sanitarias \cite{CCAES2020}.}

\section{\si2r2 model}\label{SARSmod}

This model, which can be regarded as a  variant of the SEIRD model in \cite{CDPFPG2020} (see also \cite{wiki}), subdivides the population into the following categories:

\begin{itemize}
\item {\em susceptible} people, which may be infected by the disease;
\item {\em infectious} people, {\em not yet diagnosed}, which may spread the disease;
\item {\em infectious diagnosed} people, which have been quarantined;
\item {\em removed undiagnosed} people by spontaneous recovery, which are no more infectious;
\item {\em removed diagnosed} people, either by recovery or death.
\end{itemize}
 The acronym  \si2r2 derives by the initials of the above five classes.
Moreover, we take into account that infectious people are  diagnosed with a delay,  due to  the incubation period that precedes the appearance of  symptoms (see, e.g., \cite{He2020}) and a further amount of time before the infected people are actually tested. As in the case of the SIR model, it is assumed that recovered people acquire some immunity (or are dead), so that they are no more susceptible. The model is as follows:

\begin{eqnarray}\nonumber
\dot x(t) &=& -\beta x(t) y_1(t),\\ \nonumber
\dot y_1(t) &=& \beta x(t) y_1(t) -\sigma y_1(t-\tau) s_+(y_1(t))-\gamma_1 y_1(t),\\ \label{sars}
\dot y_2(t) &=& \sigma y_1(t-\tau)s_+(y_1(t)) -\gamma_2 y_2(t),\\ \nonumber
\dot z_1(t) &=& \gamma_1 y_1(t), \\ \nonumber
\dot z_2(t) &=& \gamma_2 y_2(t),
\end{eqnarray}
where:
\begin{itemize}
\item $x(t)$ is the number of {\em susceptible} people at time $t$,
\item $y_1(t)$ is the number of {\em infectious undiagnosed} people  at time $t$,
\item $y_2(t)$ is the number of {\em infectious diagnosed} people at time $t$,
\item $z_1(t)$ is the number of {\em removed undiagnosed} infectious people (by spontaneous recovery) at time $t$,
\item $z_2(t)$ is the number of {\em removed diagnosed} infectious people (by recovery or death) at time $t$,
\end{itemize}
and
\begin{itemize}
\item $\beta$ is the {\em coefficient of infection},
\item $\sigma$ is the {\em coefficient of transition to the illness},
\item $\gamma_i$, $i=1,2$, are the {\em coefficients of removal} of undiagnosed and diagnosed infectious people, respectively,
\item $\tau$ is a {\em delay time}, occurring between the beginning of infectiousness and the diagnosys,%\footnote{A reasonable value for $\tau$ seems to be about 10 days.}
\item finally, 
\begin{equation}\label{splus}
s_+(y) = \left\{\begin{array}{lcl} 1, & &\mbox{if ~$y>0$,}\\[2mm] 0, &&\mbox{otherwise.}\end{array}\right.
\end{equation} 
\end{itemize}
As is clear from the third equation in (\ref{sars}), a basic assumption in the model is that only infectious undiagnosed  people can spread the disease, whereas infectious diagnosed people are quarantined, in some way. 

\begin{rem}\label{tau}
In the sequel, for sake of brevity, we shall use the notation $y_1^{(\tau)}(t)$ to denote $y_1(t-\tau)$. For the same reason, the argument $t$ will be often omitted.
\end{rem}
 
From (\ref{sars}), by considering that the sum of the right-hand sides is identically zero, one easily realizes that:
\begin{equation}\label{cost_1}
S(t) = x(t)+y_1(t)+y_2(t)+z_1(t)+z_2(t) \equiv const,
\end{equation}
a conservation property similar to (\ref{cost}) of the SIR model.
%until $y_1(t)$ eventually vanishes. 
Even though a complete analysis of the model will be done elsewhere, we also observe that, when the infectious diagnosed people reach their ``peak'' (where $\dot y_2 = 0$) with a value $y_2^*$, one has that the level of infectious undiagnosed people, $\tau$ days before, was given by:
$$%\begin{equation}\label{peakas}
y_1^* = \frac{\gamma_2}\sigma y_2^*.
$$%\end{equation}
Moreover, as a rough approximation, one may assume, for example, $\gamma_1\approx\gamma_2$, unless more refined estimates are available for $\gamma_1$ (an estimate of $\gamma_2$ can be usually derived from the data).

As in the case of the SIR model, (\ref{sars}) may be not appropriate when the spread of the epidemic is not homogeneous in different regions of a country, as is the case for the COVID-19 epidemic in Italy and Spain \cite{PC2020,CCAES2020}.  For this reason, in the next section we propose a  multi-region generalization of the basic \si2r2 model (\ref{sars}).

\section{Multi-region \si2r2 (mr\si2r2) model}\label{mrSARSmod}
In order to generalize the model (\ref{sars}) to a scenario where there are $n$ regions, with a different situation of the epidemic, we proceed in a way similar to what done for the SIR model in Section~\ref{mrSIRmod}. Let us then  define: 
\begin{equation}\label{XYkZk}
\bfx = \pmatrix{c} x_1\\ \vdots \\ x_n\endpmatrix, \qquad 
\bfy_k = \pmatrix{c} y_{1k}\\ \vdots \\ y_{nk}\endpmatrix, \qquad 
\bfz_k = \pmatrix{c} z_{1k}\\ \vdots \\ z_{nk}\endpmatrix, \qquad k=1,2,
\end{equation}
as the vectors with susceptible, undiagnosed infectious, diagnosed infectious, undiagnosed removed, and diagnosed removed people, respectively, in each region. I.e.:
\begin{itemize}
\item $x_i$ is the level of susceptible people in region $i$;
\item $y_{i1}$ is the level of undiagnosed infectious people in region $i$;
\item $y_{i2}$ is the level of diagnosed infectious people in region $i$;
\item $z_{i1}$ is the level of removed undiagnosed infectious people in region $i$;
\item $z_{i2}$ is the level of removed diagnosed infectious people in region $i$.
\end{itemize} 
In so doing, the basic model (\ref{sars}) becomes:
\begin{eqnarray}\nonumber
\dot \bfx &=& -B\, \bfx \circ \bfy_1 + R\bfx,\\  \nonumber
\dot \bfy_1 &=& B\, \bfx \circ \bfy_1 -\Sigma\bfy_1^{(\tau)}\circ s_+(\bfy_1) -C_1 \bfy_1 + R\bfy_1,\\ \label{mrsars}
\dot \bfy_2 &=&  \Sigma\bfy_1^{(\tau)}\circ s_+(\bfy_1)-C_2 \bfy_2,\\ \nonumber
\dot \bfz_1 &=& C_1 \bfy_1+R\bfz_1, \\  \nonumber
\dot \bfz_2 &=& C_2 \bfy_2,
\end{eqnarray}
where\,\footnote{As before, $\circ$ denotes the Hadamard (i.e., component-wise) product.} 
$$\bfy_1^{(\tau)}=\pmatrix{c}y_{11}^{(\tau)}\\ \vdots \\ y_{n1}^{(\tau)}\endpmatrix, \qquad
s_+(\bfy_1)=\pmatrix{c}s_+(y_{11})\\ \vdots \\ s_+(y_{n1})\endpmatrix,$$  
are defined according to Remark~\ref{tau} and Equation (\ref{splus}), respectively, matrices $B$ and $R$ are formally defined as in (\ref{BCR}), and
\begin{equation}\label{CS}
\Sigma = \pmatrix{ccc} \sigma_1\\ &\ddots\\ &&\sigma_n\endpmatrix,\qquad
C_k = \pmatrix{ccc} \gamma_{1k}\\ &\ddots\\ &&\gamma_{nk}\endpmatrix,\quad k=1,2, 
\end{equation}
are the matrices with the transition to illness, and removal coefficients (of undiagnosed and diagnosed infectious, respectively) in each region.  As in the case of the mrSIR model, we assume that the entries of the migration matrix $R$ satisfy (\ref{R})-(\ref{R1}), 
thus providing the conservation property
\begin{equation}\label{cost1_1}
S(t) = \sum_{i=1}^n \left[x_i(t)+y_{i1}(t)+y_{i2}(t)+z_{i1}(t)+z_{i2}(t)\right] \equiv const.
\end{equation}
Clearly, (\ref{cost1_1}) is the analogous of (\ref{cost_1}) for the multi-region version of the model.

\begin{rem}
As is clear from (\ref{mrsars}), we have allowed migration for susceptible, undiagnosed infectious, and recovered undiagnosed people only. However, it is straightforward to generalize migration to recovered diagnosed people, too, by distinguishing between {\em cured} and {\em death} people.
\end{rem}
 In componentwise form, (\ref{mrsars}) reads, by taking into account (\ref{R})-(\ref{R1}):
\begin{eqnarray}\nonumber
\dot x_i &=& -\beta_i x_i y_{i1} + \sum_{j=1}^n\rho_{ij} x_j,\\   \nonumber
\dot y_{i1} &=& \beta_i x_i y_{i1} - \sigma_i y_{i1}^{(\tau)}s_+(y_{i1}) - \gamma_{i1} y_{i1} + \sum_{j=1}^n\rho_{ij} y_{j1},\\ \label{mrsarsi}
\dot y_{i2} &=& \sigma_i y_{i1}^{(\tau)}s_+(y_{i1}) - \gamma_{i2} y_{i2},\\[2mm] \nonumber
\dot z_{i1} &=& \gamma_{i1}y_{i1} + \sum_{j=1}^n\rho_{ij} z_{j1},\\[1mm] \nonumber
\dot z_{i2} &=& \gamma_{i2} y_{i2},  \qquad\qquad\qquad\qquad i=1,\dots,n.
\end{eqnarray}
\begin{rem} As in the single region case, as an approximation in (\ref{mrsarsi}) one may assume $$\gamma_{i1}\approx \gamma_{i2},\qquad  i=1,\dots,n,$$ (i.e., $C_1\approx C_2$ in (\ref{mrsars})), in order to simplify the use of the model, unless better estimates are available.
\end{rem}

\bigskip

We conclude this note, by observing that both the mrSIR  and the mr\si2r2 models can be used at different space scales, i.e.: regions within a country; countries within a continent or worldwide; etc.

\end{document}